\documentclass[preprint,showpacs,preprintnumbers,amsmath,amssymb,aps]{revtex4}

\usepackage{graphicx}
\usepackage{dcolumn}
\usepackage{bm}

\begin{document}

\preprint{PRESAT-8002}

\title{First-Principles Study on Electron-Conduction Properties of C$_{60}$ Chains}

\author{Tomoya Ono and Kikuji Hirose }
\affiliation{Department of Precision Science and Technology, Osaka University, Suita, Osaka 565-0871, Japan }

\date{\today}

\begin{abstract}
The electron-conduction properties of fullerene chains are examined by first-principles calculations based on the density functional theory.  The conductivity of the C$_{60}$ dimer is low owing to the constraint of the junction of the molecules on electron conduction, whereas the C$_{60}$ monomer exhibits a conductance of $\sim$ 1 G$_0$. One of the three degenerate $t_{u1}$ states of C$_{60}$ is relevant to conduction and the contributions of the others are small. In addition, we found a more interesting result that the conductance of the fullerene chain is drastically increased by encapsuling metal atoms into cages.
\end{abstract}

\maketitle
Recently, extensive research has been devoted to understanding electron conduction in atomic- and molecular-scale devices. In particular, carbon-based molecules, such as fullerenes and carbon nanotubes, have attracted considerable attention, because they are expected to play a key role in nanoscale electronic devices owing to their mechanical stability. Thus, their electronic and mechanical properties have been intensively studied \cite{fpfullerene}. In the preceding study \cite{otani}, we examined the electron-conduction properties of C$_{20}$ chains and revealed that the conductances of monomers are $\sim$ 1 G$_0$ while those of dimers are reduced to $\sim$ 0.2 G$_0$, in which most of the incident electrons are scattered at the junctions between the molecules in the case of the dimers. Although C$_{20}$ chains are expected to be more suitable for extremely precise molecular electronic devices than carbon nanotubes or other fullerenes, the atomic structures of C$_{20}$ chains are not easily controllable at present. In the last decade, some theoretical studies have reported that a single C$_{60}$ bridge exhibits metallic properties and that its conductance is between 1 and 3 G$_0$ \cite{palacios}. In addition, some exciting experiments demonstrate that C$_{60}$ molecules have the possibility of being a nanoscale electrical amplifier \cite{joachim}. Moreover, several experiments have reported that polymeric forms of C$_{60}$ are synthesized under special conditions \cite{polymer} (high pressure, photoexcitation, or in the charged/doped phase), and endohedral metal fullerenes, in which metal atoms are encapsuled in cages, are produced during the collisions of metal ions with C$_{60}$ vapor molecules \cite{endhedral}. These fullerenes are good candidates for nanoscale devices, yet there remains much to be learned about their electron-conduction properties.

In this Letter, we examined the electron conduction through C$_{60}$ chains suspended between semi-infinite gold electrodes using first-principles calculations within the framework of the density functional theory \cite{dft}. We found that the conductance of the C$_{60}$ monomer is $\sim$ 1 G$_0$, while that of the C$_{60}$ dimer is $\sim$ 0.1 G$_0$ owing to the scattering of incident electrons at the junction between the molecules. One of the three degenerate $t_{u1}$ states is dominant in electron conduction, whereas the others scarcely contribute to electron conduction. Moreover, when lithium atoms are encapsuled in C$_{60}$ cages, the energy of the unoccupied molecular orbitals at the junction shifts down to the Fermi level, and as a consequence, the conductance of the Li@C$_{60}$ dimer significantly increases.

\begin{figure}[htb]
\begin{center}
\includegraphics{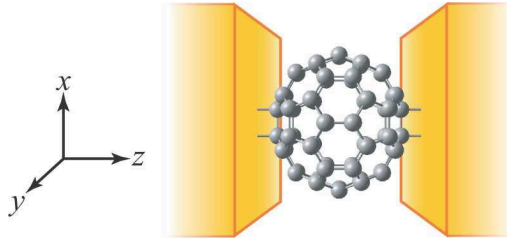}
\end{center}
\caption{(color online) Schematic diagram of scattering region of C$_{60}$ molecules suspended between Au jellium electrodes.}
\label{fig:fig1}
\end{figure}

Our first-principles calculation method for electron-conduction properties is based on the real-space finite-difference approach \cite{rsfd,icp,tsdg}, which enables us to determine the self-consistent electronic ground state with a high degree of accuracy by the timesaving double-grid technique \cite{icp,tsdg} and the direct minimization of the energy functional \cite{dm}. Moreover, the real-space calculations eliminate the serious drawbacks of the conventional plane-wave approach, such as its inability to describe nonperiodic systems accurately. The norm-conserving pseudopotentials \cite{norm} of Troullier and Martins \cite{tmpp} are employed to describe the electron-ion interaction, and exchange correlation effects are treated by the local density approximation \cite{lda} of the density functional theory \cite{dft}.

\begin{figure}[htb]
\begin{center}
\includegraphics{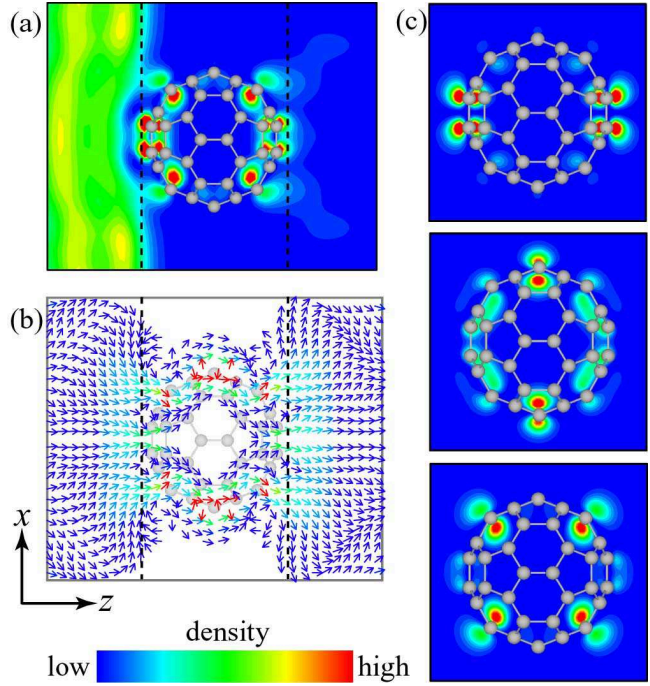}
\end{center}
\caption{(color) (a) Current charge distribution of incident electrons from left electrode at the Fermi level when C$_{60}$ is suspended between electrodes. (b) Current distribution of incident electrons. The planes shown are perpendicular to the electrode surfaces and contain the atoms facing the electrodes. (c) Charge distributions of three degenerate LUMOs of isolated C$_{60}$ molecule for $t_{u1,x}$ (upper panel), $t_{u1,y}$ (middle panel) and  $t_{u1,z}$ (lower panel). The spheres, lines, and vertical dashed lines represent carbon atoms, C-C bonds, and the edges of the jellium electrodes, respectively.}
\label{fig:fig2}
\end{figure}

\begin{figure}[htb]
\begin{center}
\includegraphics{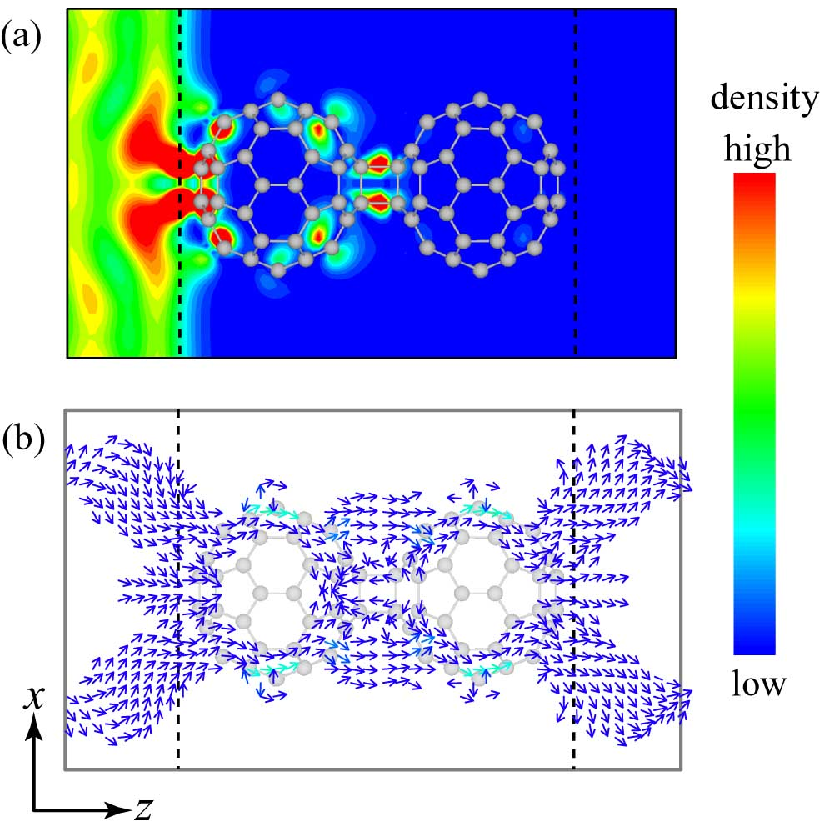}
\end{center}
\caption{(color) (a) Current charge distribution of incident electrons from left electrode at the Fermi level when (C$_{60}$)$_2$ is suspended between electrodes. (b) Current distribution of incident electrons. The planes shown and the meanings of the symbols are the same as in Fig.~\ref{fig:fig2}.}
\label{fig:fig3}
\end{figure}

Figure \ref{fig:fig1} shows the calculation model employed here, in which the C$_{60}$ molecule is sandwiched between electrodes. Because polymerized C$_{60}$ was found to form a parallel double-bonded structure in the previous study \cite{polydimer}, a model in which a C$_{60}$ molecule is connected to electrodes with parallel double bonds is employed. Since many first-principles investigations using structureless jellium electrodes are in almost quantitative agreement with experiments, we substitute the jellium electrodes for crystal ones. To determine the optimized atomic structures and Kohn-Sham effective potential, a conventional supercell is employed under the periodic boundary condition in all directions; the size of the supercell is $L_x=L_y=30.8$ a.u. and $L_z=L_{mol}+25$ a.u., where $L_x$ and $L_y$ are the lateral lengths of the supercell in the $x$- and $y$-directions parallel to the electrode surfaces, respectively, $L_z$ is the length in the $z$-direction, and $L_{mol}$ is the length of the inserted molecules. The C$_{60}$ monomers and dimers are put between the electrodes. The structural optimizations are implemented in advance with a cutoff energy of 89 Ry, which corresponds to a real-space grid spacing of $\sim$ 0.33 a.u., and a higher cutoff energy of 799 Ry in the vicinity of nuclei with the augmentation of the double-grid technique \cite{icp,tsdg}. The molecules are optimized individually under the isolated condition and then relaxed between the electrodes. Although the initial distance between the edge atoms of inserted molecules and the jellium electrode is set at 0.91 a.u. \cite{comment}, this distance increases after the structural optimization of the molecules with the jellium electrodes. For the conductance calculation, we take a cutoff energy of 39 Ry. The Kohn-Sham effective potential is computed using the supercell employed in the structural optimization. We ensured that the increase in the cutoff energy and the enlargement of the supercell did not affect our conclusion significantly. The global wave functions for infinitely extended states from one electrode to the other are evaluated by the overbridging boundary-matching formula \cite{icp,obm}. The conductance of the nanowire system at the limits of zero temperature and zero bias is described by the Landauer-B\"uttiker formula, G=tr({\bf T}$^\dag${\bf T}) G$_0$ \cite{buttiker}, where {\bf T} is the transmission matrix. To investigate states actually contributing to electron conduction, the eigenchannels are computed by diagonalizing the Hermitian matrix {\bf T}$^\dag${\bf T} \cite{nkobayashi} and their charge distributions are compared with those of the orbitals of the isolated molecule.

\begin{figure}[htb]
\begin{center}
\includegraphics{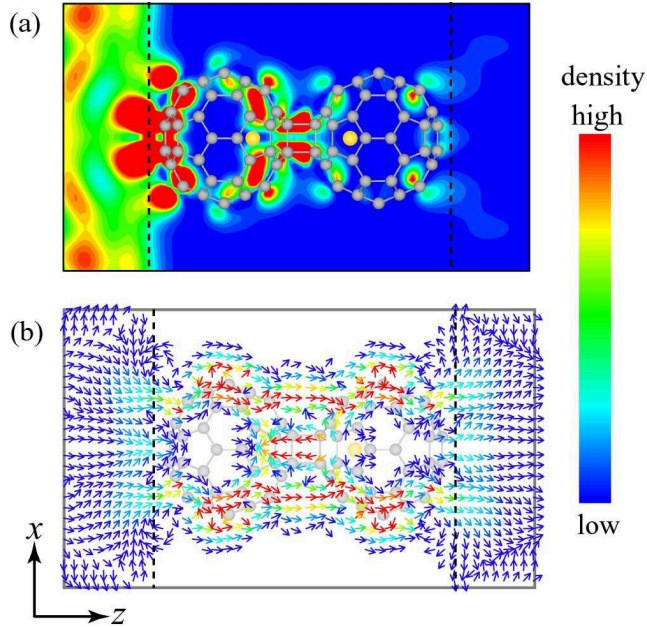}
\end{center}
\caption{(color) (a) Current charge distribution of incident electrons from left electrode at the Fermi level when (Li@C$_{60}$)$_2$ is suspended between electrodes. (b) Current distribution of incident electrons. The planes shown are the same as in Fig.~\ref{fig:fig2}. The yellow spheres are lithium atoms and the meanings of the other symbols are the same as in Fig.~\ref{fig:fig2}.}
\label{fig:fig4}
\end{figure}

We first calculate the electron-conduction properties of the C$_{60}$ monomer. Figure~\ref{fig:fig2} shows the current charge distribution and current distribution of the incident electrons from the left electrode at the Fermi level, where the cross sections at the center of the models along the $z$-direction are depicted. The electron currents pass along the C-C bonds, and not directly cross through the inside of the C$_{60}$ cages. The charge distributions of the lowest unoccupied molecular orbitals (LUMOs) of the isolated C$_{60}$ monomer are shown in Fig.~\ref{fig:fig2}(c). The C$_{60}$ has three $t_{u1}$ degenerate LUMOs and we name, for convenience, the states that are odd functions with respect to the $y$-$z$, $z$-$x$ and $x$-$y$ planes as $t_{u1,x}$, $t_{u1,y}$ and $t_{u1,z}$, respectively. The incident electrons are found to be scattered at both interfaces between the molecule and the electrodes. The channel transmissions of the respective states are collected in the second column of Table~\ref{tbl:tbl1}. It can be seen that the $t_{u1,z}$ state dominantly contributes to electron conduction. According to the wave-function matching formula for the three-dimensional jellium wire \cite{egami}, the transmission of the channel that has high kinetic energy in the direction parallel to the chain is large and is hardly affected by the length of the wire. The transmission of the $t_{u1,z}$ state is higher than those of the other degenerate $t_{u1}$ states, since its kinetic energy in the $z$-direction is the largest. The resultant conductance is 1.13 G$_0$, which is in agreement with the previous theoretical calculations, which report that the C$_{60}$ monomer exhibits metallic electron-conduction properties and that its conductance is between 1 and 3 G$_0$ \cite{palacios}.

\begin{table}[thbp]
\begin{center}
\caption{Channel transmissions at the Fermi level.}
\label{tbl:tbl1}

\begin{tabular}{c|ccc} \hline\hline
              & C$_{60}$ & (C$_{60}$)$_2$ & (Li@C$_{60}$)$_2$ \\ \hline
$t_{u1,x}$    & 0.110   &      0.001    &     0.007     \\
$t_{u1,y}$    & 0.133   &      0.012    &     0.020     \\
$t_{u1,z}$    & 0.873   &      0.098    &     0.851     \\ \hline\hline
\end{tabular}
\\
\end{center}
\end{table}

\begin{figure}[htb]
\begin{center}
\includegraphics{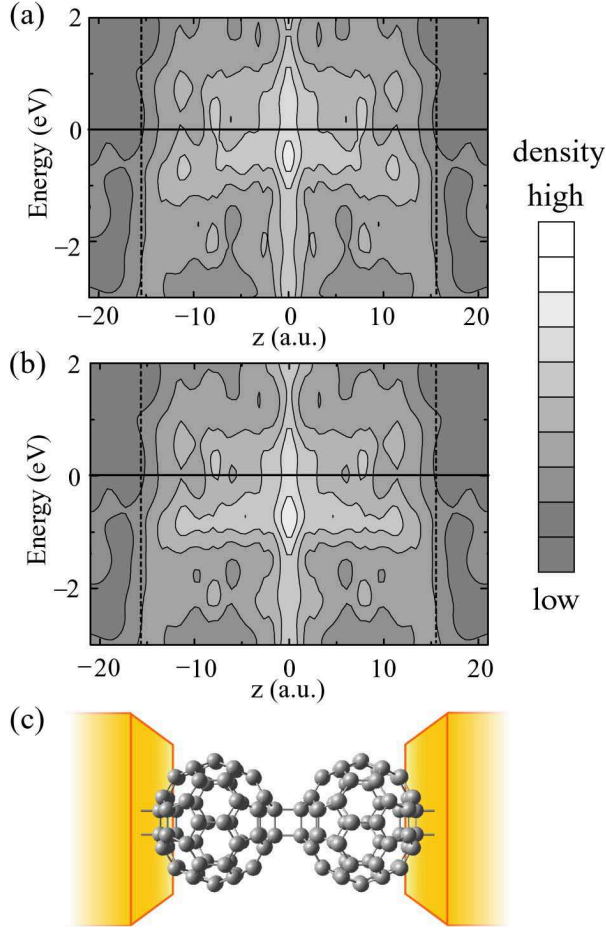}
\end{center}
\caption{(color online) Distributions of DOS integrated on plane perpendicular to chain as functions of relative energy from the Fermi level for (a) (C$_{60}$)$_2$ chain and (b) (Li@C$_{60}$)$_2$ chain. The zero of energy is chosen to be the Fermi level. Each contour represents twice or half the density of adjacent contour lines. The vertical dashed lines represent the edge of the jellium electrodes. A schematic diagram of the computational model for the C$_{60}$ dimer is illustrated in (c) as a visual aid.}
\label{fig:fig5}
\end{figure}

Next, the conduction property of the dimer is examined. Figure~\ref{fig:fig3} shows the current charge distribution and current distribution at the Fermi level. Since the reflection of electrons occurs at the junction in addition to the interfaces, the conductance has a lower value (0.11 G$_0$) than that of the C$_{60}$ monomer. This situation is similar to that of C$_{20}$ \cite{otani}. The channel transmissions are listed in the third column of Table~\ref{tbl:tbl1}. Note that all the channel transmissions are lower than those of C$_{60}$. In the case of the isolated C$_{60}$ dimer, the characteristics of the first and second LUMOs of the isolated dimer are inherited from the $t_{u1,y}$ and $t_{u1,z}$ LUMOs of the isolated C$_{60}$ molecule, respectively. The contribution of the second LUMO to electron conduction is the largest, although its energy is higher than that of the first LUMO. According to our result, the conductances of the longer chains are expected to decay rapidly as the number of fullerenes in the chains increases because of their nonmetallic properties. Do fullerene-based chains never work as conductive molecule wires?

It has been reported that endohedral metal fullerenes, in which metal atoms are encapsulated in cages, are produced by the collisions of metal ions with C$_{60}$ vapor molecules \cite{endhedral}. Since the volume of C$_{60}$ is larger than that of C$_{20}$, we can insert metal atoms into the cages. Here, lithium atoms are inserted into the cages so as to locate on the central axis of the chain. The total energy gain realized by encapsuling the lithium atoms is 3.89 eV per dimer, the cohesive energy of the (Li@C$_{60}$)$_2$ molecule is 0.10 eV per Li@C$_{60}$ molecule, and the energy difference between the highest occupied molecular orbital and LUMO is 0.21 eV, which is markedly smaller than that of (C$_{60}$)$_2$ (1.33 eV). We show in Fig.~\ref{fig:fig4} the current charge distribution and current distribution of the (Li@C$_{60}$)$_2$ molecule. The reflection at the junction of the molecules is suppressed and the conductance reaches 0.88 G$_0$. We show the channel transmissions in the fourth column of Table~\ref{tbl:tbl1}. The transmissions of the states consisting of the $t_{u1,z}$ orbital are significantly larger than those of (C$_{60}$)$_2$ and recover up to that of the monomer. To interpret the reason for the increase in conductance, Fig.~ \ref{fig:fig5} shows the density of states (DOS) of the  (C$_{60}$)$_2$ and  (Li@C$_{60}$)$_2$ suspended between the electrodes, which are plotted by integrating the DOS on the plane perpendicular to the chains. The encapsuled lithium atoms do not largely affect the electronic structure at the interface. Yet, the states that are $\sim$ 0.3 eV above the Fermi level of the C$_{60}$ dimer is shifted down to the Fermi level by inserting the lithium atoms, which lead to an increase in the DOS at the Fermi level in the fullerenes. The low energy gap between the Fermi level and the unoccupied states at the junction, and the high DOS at the Fermi level result in small reflection at the junction and the high conductance of the endohedral metal fullerene chain.

In conclusion, we have theoretically investigated the electron-conduction properties of C$_{60}$ chains. The $t_{u1,z}$ state is closely relevant to electron conduction, while the others hardly contribute to conduction. The C$_{60}$ dimer does not possess good conductivity, because the junction of the molecules is a bottleneck of electron conduction. On the other hand, by inserting the lithium atoms into the cages, the C$_{60}$ dimer exhibits good conduction properties due to the increase in the DOS at the junction. These results indicate that the endohedral metal fullerene chains have potential applications in electronic devices. In addition, encapsuling metal atoms into fullerene will be a novel scheme of controlling the electron-conduction properties of the fullerene chains. Thus, this situation will stimulate a new technology of electronic devices using molecule chains.

This research was partially supported by a Grant-in-Aid for the 21st Century COE ``Center for Atomistic Fabrication Technology'', by a Grant-in-Aid for Scientific Research in Priority Areas ``Development of New Quantum Simulators and Quantum Design'' (Grant No. 17064012) and also by a Grant-in-Aid for Young Scientists (B) (Grant No. 17710074) from the Ministry of Education, Culture, Sports, Science and Technology. The numerical calculation was carried out by the computer facilities at the Institute for Solid State Physics at the University of Tokyo and the Information Synergy Center at Tohoku University.

\end{document}